%% file: 2019Mandurrino_CTDWIT.tex
\documentclass[10pt, paper=a4, UKenglish]{article}
\usepackage{graphicx}
\def\Title#1{\begin{center} {\Large #1 } \end{center}}
\def\Author#1{\begin{center}{ \sc #1} \end{center}}
\def\Address#1{\begin{center}{ \it #1} \end{center}}

\newcommand\pubblock{\rightline{\begin{tabular}{l} Proceedings of the CTD/WIT 2019\\ \pubnumber\\
         \pubdate  \end{tabular}}}

\newenvironment{Abstract}{\begin{quotation} \begin{center} 
             \large ABSTRACT \end{center}\bigskip 
      \begin{center}\begin{large}}{\end{large}\end{center} \end{quotation}}

\newenvironment{Presented}{\begin{quotation} \begin{center} 
             PRESENTED AT\end{center}\bigskip 
      \begin{center}\begin{large}}{\end{large}\end{center} \end{quotation}}

\input econfmacros.tex

\usepackage{color}
\usepackage{lineno}
\usepackage{subfig}
\usepackage{hyperref}
\usepackage{amsmath}

\newcommand\pubnumber{PROC-CTD19-112}

\newcommand\pubdate{\today}

\def\affiliation{
On behalf of the RD50 Collaboration, \\
CERN, Geneva, Switzerland}

\newcommand{\conference}{Connecting the Dots and Workshop on Intelligent Trackers (CTD/WIT 2019)\\
Instituto de F\'isica Corpuscular (IFIC), Valencia, Spain\\ 
April 2-5, 2019}

\usepackage{fancyhdr}
\definecolor{mygrey}{RGB}{105,105,105}
\begin{document}


\large
\begin{titlepage}
\pubblock

\vfill
\Title{Silicon detectors for the LHC Phase-II upgrade and beyond.\\RD50 status report}
\vfill

\Author{Marco Mandurrino}
\Address{\affiliation}
\vfill

\begin{Abstract}
It is foreseen to significantly increase the luminosity of the LHC by upgrading towards the \mbox{HL-LHC} (\mbox{High-Luminosity} LHC) in order to harvest the maximum physics potential. Especially the \mbox{Phase-II-Upgrade} foreseen for 2023 will mean unprecedented radiation levels, significantly beyond the limits of the Silicon trackers currently employed. All-Silicon central trackers are being studied in ATLAS, CMS and LHCb, with extremely radiation-hard Silicon sensors to be employed on the innermost layers. Within the RD50 Collaboration, a large R\&D program has been underway for more than a decade across experimental boundaries to develop Silicon sensors with sufficient radiation tolerance for \mbox{HL-LHC} trackers. Key areas of recent RD50 research include new sensor fabrication technologies such as \mbox{High-Voltage} \mbox{(HV)-CMOS}, exploiting the wide availability of the CMOS process in the semiconductor industry at very competitive prices compared to the highly specialized foundries that normally produce particle detectors on small wafers. We also seek for a deeper understanding of the connection between the macroscopic sensor properties such as radiation-induced increase of leakage current, doping concentration and trapping, and the microscopic properties at the defect level. Another strong activity is the development of advanced sensor types like 3D Silicon detectors, designed for the extreme radiation levels expected for the vertexing layers at the \mbox{HL-LHC}. A further focus area is the field of \mbox{Low-Gain} Avalanche Detectors (LGADs), where a dedicated multiplication layer to create a high field region is built into the sensor. LGADs are characterized by a high signal also after irradiation and a very fast signal compared to traditional Silicon detectors with make them ideal candidates for ATLAS and CMS timing layers in the \mbox{HL-LHC}. We will present the state of the art in several Silicon detector technologies as outlined above and at radiation levels corresponding to \mbox{HL-LHC} fluences and partially beyond.
\end{Abstract}

\vfill

\begin{Presented}
\conference
\end{Presented}
\vfill
\end{titlepage}
\def\thefootnote{\fnsymbol{footnote}}
\setcounter{footnote}{0}
%

\normalsize 


\section{Introduction}
\label{intro}

The RD50 Collaboration is a CERN community devoted to the development of \mbox{radiation-hard} semiconductor devices for \mbox{very-high-luminosity} colliders which brings together experts from more than 60 institutes all over the world, such as solid-state physicists, device physicists, experts of \mbox{radiation-matter} interactions, high energy physicists, electronics system designers, ASICs designers and sensor foundries. This community has learned to work towards common goals, developing methods, tools and standards that are world reference for the domain. Thanks to this vast expertise, RD50 put the unsurpassed basis of the R\&D towards sensor solution for future challenges in several fields. To this aim, the Collaboration has been organized in four complementary divisions, working in parallel and synergistically, in a sort of \mbox{bottom-up} approach: we have the characterization of materials and their defects (due to production processes or radiation) and, due to the capability to design and produce devices, we are also devoted to the characterization of detectors, then we find the realization of new structures and, finally, full detector systems. The microscopic analysis of Silicon and other materials through their standard, \mbox{defect-engineered} properties, both pre- and \mbox{post-irradiation}, is essential to the refinement of several physics-based models involved in the simulation and design of particle detectors. By characterizing the test structures, through large data sets of $I(V)$, $C(V)$ and charge collection curves, after irradiation (up to fluences where this is still relevant) and annealing, we parametrize their evolution and optimize such models. Among the main RD50 achievements from 2002 we have the development of the \mbox{$p$-type} Silicon strip and pixel technology, as well as LGAD (\mbox{Low-Gain} Avalanche Diodes), double column 3D detectors and the demonstration of the performance of planar segmented sensors to the maximum fluences anticipated for the \mbox{HL-LHC} (\mbox{$3\cdot10^{16}~\mathrm{n}_\mathrm{eq}/\mathrm{cm}^2$}). Also noticeable is the extensive evaluation of defect engineered Silicon and other semiconductor materials and characterization, with the identification of defects responsible for the degradation of various detectors \mbox{figure-of-merit} defining the \mbox{state-of-the-art} in the corresponding \mbox{solid-state} community. Moreover, we developed several unique characterization methods and systems for sensor and material analyses, such as the Transient Current Technique (TCT)~\cite{1993Eremin_NIMA}, \mbox{edge-TCT}, \mbox{Two-Photon} \mbox{Absorption-TCT} (\mbox{TPA-TCT}), the ALiBaVa readout system, besides many other standardized measurement and analyses procedures which now are partly marketed through \mbox{spin-off} companies. Finally, RD50 also boasts the data collection and development of damage parameters/models essential for sensor design (for instance, as TCAD input data) and for planning the scenarios of future HEP experiments and their upgrades, in very close links to the current LHC experiments and their upgrades.

\section{Radiation damage and defects characterization}
\label{section2}

\begin{figure}[!htb]
\centering
\includegraphics[width=0.5\linewidth]{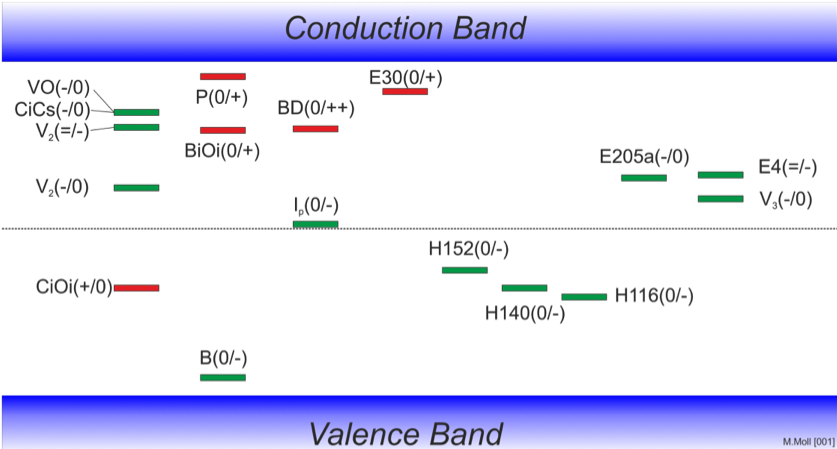}
\caption{Picture of the most important defect levels populating the energy bandgap in Silicon (along with the indication of their typical charge). Levels P(0/+), close to the conduction band, and B(0/-), just above the valence band energy, correspond to the $n$- and $p$-type dopants, respectively.}
\label{fig:defects_map}
\end{figure}

The main goals of the activities focusing on material and defects characterization are the identification of defects responsible for trapping, leakage current, change of charge collection efficiency (CCE), effective doping concentrations or electric field, the evaluation of any possible positive implication of such knowledge to mitigate radiation damage effects and the delivery of input device simulations to predict detector performance under various conditions, also by providing irradiated samples. As a result of all these activities, a \mbox{full-comprehensive} energy map of the most relevant defects in Silicon has been set up (see Figure~\ref{fig:defects_map}), allowing noticeable progresses in the knowledge about the occurrence of increased generation or leakage currents, trapping or CCE degradation. The specific analysis which allowed such step forward in characterizing defects are several and include \mbox{capacitance-Deep-Level} Transient Spectroscopy (\mbox{c-DLTS}), Thermally Stimulated Currents (TSC), \mbox{Photo-Induced} Transient Spectroscopy (PITS), Fourier Transform Infrared Spectroscopy (FTIR) and the already mentioned TCT.

In Figure~\ref{fig:defects_map} one may see the most important defect levels in Silicon, which is the converging result of a consistent set of data observed after $\pi$, neutron, proton, photon and electron irradiation campaigns. The introduction rate of such defects is depending on particle type, energy and, to some extent, on materials. Particularly noticeable is the indication that E205a and H125 are important in the trapping process, while some deep acceptors (like H116) and shallow donors (e.g. E30) may alter the space charge. Moreover, there are several evidences that some defects, only seen in \mbox{$p$-type} (Boron containing) materials, participate to the acceptor removal (or \mbox{de-activation}) mechanism involving \mbox{B$_\mathrm{i}$O$_\mathrm{i}$} complexes.

\begin{figure}[!htb]
\centering
\includegraphics[width=0.5\linewidth]{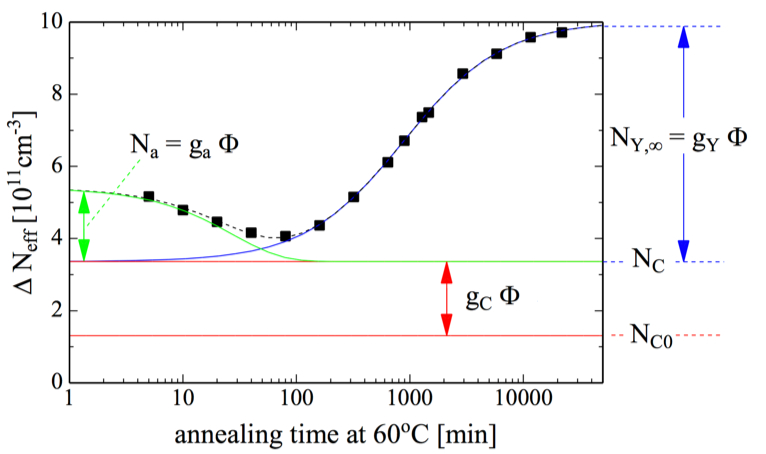}
\caption{Trend of $\Delta N_\mathrm{eff}(\phi,t) = N_\mathrm{eff}(0) - N_\mathrm{eff}(\phi,t)$ versus the annealing time $t$, whose parametrization is known as the Hamburg model.}
\label{fig:Hamburg_model}
\end{figure}

In the following we briefly review some relevant RD50 activities belonging to the field of material characterization. The first one is the well-known Hamburg model, a milestone in the knowledge of defects properties and their behavior as a function of fluence $\phi$ (in units of neutrons equivalent per cm$^2$) and annealing time $t$. According to the model, which has been extensively validated with laboratory measurements for different Silicon substrates and experimental setup, the effective doping concentration is subject to a perturbation, whose magnitude
\begin{linenomath*}
\postdisplaypenalty=0
\begin{align}
\Delta N_\mathrm{eff}(\phi,t) &= N_\mathrm{eff}(0) - N_\mathrm{eff}(\phi,t)\nonumber\\
&= N_\mathrm{a}(\phi,t) + N_\mathrm{C}(\phi) + N_\mathrm{Y}(\phi,t)
\end{align}
\end{linenomath*}
is a function of three different terms: (\textit{i}) the short-term annealing
\begin{linenomath*}
\begin{equation}
N_\mathrm{a}(\phi,t) = \phi \, \sum_i \left[ g_{\mathrm{a},i} \cdot \mathrm{exp}\left( -t/ \tau_i \right) \right] ,
\end{equation}
\end{linenomath*}
which describes a \mbox{first-order} decay of acceptors, proportional to the fluence, (\textit{ii}) the stable damage
\begin{linenomath*}
\begin{equation}
N_\mathrm{C}(\phi) = N_\mathrm{C,0} \cdot \left[ \mathrm{exp}\left(-c \cdot \phi \right)\right] + g_\mathrm{C} \cdot \phi \,,
\end{equation}
\end{linenomath*}
modeling the donor \mbox{de-activation} and the introduction of stable \mbox{acceptor-like} defects, then (\textit{iii}) the \mbox{long-term} reverse annealing contribution
\begin{linenomath*}
\begin{equation}
N_\mathrm{Y}(\phi,t) = N_\mathrm{Y,\infty}(\phi) \cdot \left(1-\frac{\tau}{1+t}\right) ,
\end{equation}
\end{linenomath*}
with $N_\mathrm{Y,\infty}(\phi) = g_\mathrm{Y} \cdot \phi$, a \mbox{second-order} parametrization, independent of fluence. The terms $g_\mathrm{a}$, $g_\mathrm{C}$ and $g_\mathrm{Y}$ are constant, and are indicated in Figure~\ref{fig:Hamburg_model}.

The Hamburg model represents the perfect example of successful interaction between data acquisition, parametric analysis and physical modeling, which is demonstrated by its vast use over the entire community working on Silicon detectors.

The second characterization activity we present consists in the \mbox{so-called} NitroStrip, a project working on \mbox{nitrogen-enriched} Silicon. As reported in several publications (see, for instance, Ref~\cite{2014Kaminski_RD50}) there is the experimental evidence that, in \mbox{$n$-type} Float Zone (FZ) substrates irradiated at \mbox{$\phi=5\cdot10^{14}$~n$_\mathrm{eq}$/cm$^2$}, the trap density is lower in samples with high nitrogen concentration. This could be due to several mechanisms reported in the literature involving the kinetics of N$_2$ complexes in Silicon, such as the vacancy suppression or the interstitial suppression~\cite{2001Ammon_JCG}.

As the last examples of ongoing activities we have to mention the very interesting results coming from mixed irradiation campaigns~\cite{2018Gosewisch_RD50}. By irradiating \mbox{magnetic-Czochralski} (MCZ) $n$-type substrates at \mbox{$\phi=3\cdot10^{14}$~n$_\mathrm{eq}$/cm$^2$} with \mbox{23~MeV} protons from Karlsruher Institut f\"{u}r Technologie (KIT), Germany, and then with neutrons from Jozef Stefan Institute (JSI) in Ljubljana at the same fluence, it has been obtained a leakage current which is larger than what observed by inverting the order of these two irradiations. Moreover, in the case of using first neutrons and then protons, the leakage current decreases after the second irradiation. Being a completely unexpected result, further analyses are needed.

\section{Detectors characterization}
\label{section3}

Since radiation-related defects may affect the detectors behavior, it is necessary to transfer all the knowledges coming from the material characterization to the device design. Not only the leakage current and other static characteristics are modified after the irradiation (as we have just seen with the NitroStrip project) but also the detector dynamics can be involved. In general, the defects introduced by the radiation may act as trapping centers for the charge carriers. As a result, the signals due to the passage of an ionizing particle can be broadened in time and then degraded, especially for timing applications. When the effective trapping time becomes of the order of the electron/hole drift time (\mbox{$\sim$5-20~ns}) the charge integrated at the electrodes by the front-end electronics is reduced. Such reduction is inversely proportional to the fluence:
\begin{linenomath*}
\postdisplaypenalty=0
\begin{align}
Q_{e,h}(\phi,t) &= Q_{e,h}^0 \cdot \mathrm{exp}\left(- \frac{t}{\tau_{\mathrm{eff}\,e,h}} \right)\nonumber\\
&= Q_{e,h}^0 \cdot \mathrm{exp}\left(- \beta_{e,h}(T,t) \cdot \phi \right) ,
\end{align}
\end{linenomath*}
where $Q_{e,h}^0$ is the electron/hole charge before irradiation and $\tau_{\mathrm{eff}\,e,h}$ the effective trapping time.

\begin{figure}[!htb]
\centering
\includegraphics[width=0.75\linewidth]{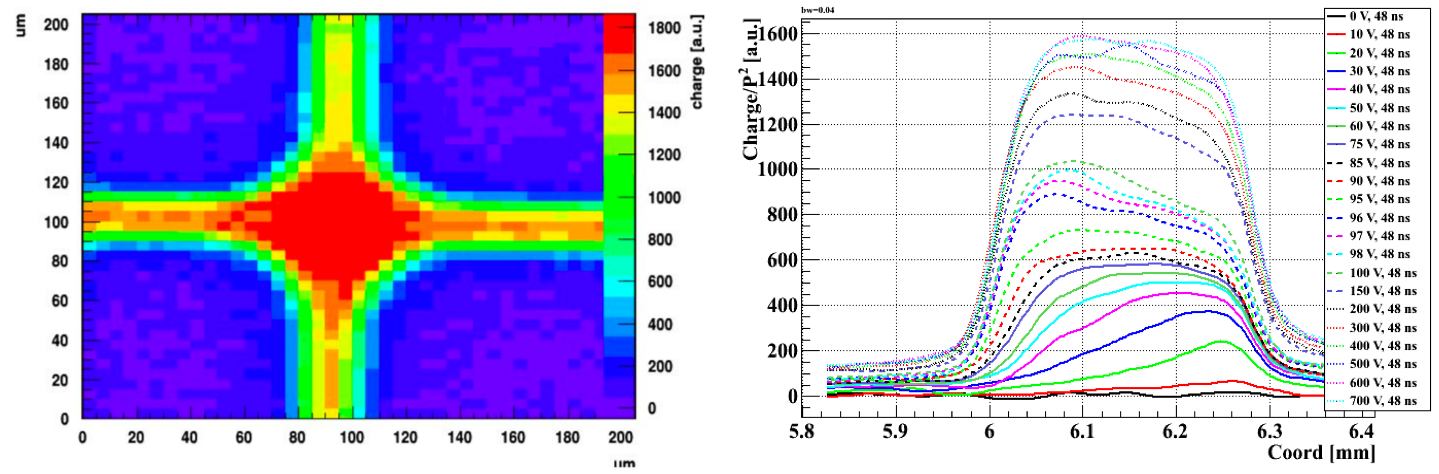}
\caption{Left: 2D map reconstructing the integrated charge in a \mbox{2$\times$2} LGAD sensor module from the UFSD3 run by FBK~\cite{2018Siviero_RD50} with the TCT setup. Right: collected charge as a function of depth in an LGAD measured with the \mbox{top-Two-Photon} Absorption technique.}
\label{fig:TCT}
\end{figure}

So, besides standard analysis of devices, like the static $I(V)$ or $C(V)$ curves, RD50 necessarily provides also dynamic characterizations. A virtuous example comes from the TCT measurements. By stimulating the detector with a pulsed IR laser beam (typically \mbox{$\lambda\sim1060$~nm}) with a \mbox{few-micron} spot size, one is able to produce the ionization of lattice atoms along all the sample thickness and, in particular, to mimic the effects of a Minimum Ionizing Particle (MIP) traveling throughout the sensor. This is especially useful to probe the performance of \mbox{Low-Gain} Avalanche Diodes (LGAD), that have segmented active areas: when the laser illuminates the implants producing charge multiplication, an amplification of the output signal is observed, and by comparing gain and no-gain signals it is possible to reconstruct the response distribution along the whole sensor and measure the width of the \mbox{inter-pixel} region, very useful to perform \mbox{fill-factor} studies. Figure~\ref{fig:TCT} shows, on the left, the 2D map of collected charges obtained with the TCT setup by integrating the signals of a \mbox{2$\times$2} LGAD from the UFSD3 production run by Fondazione Bruno Kessler (FBK), Italy.~\cite{2018Siviero_RD50}.

This is for what concerns the scan of the detector's properties along the illumination surface. All the physical quantities of interest that are functions of depth can be investigated through different methods: for instance, with the \mbox{edge-TCT} (same setup as the previous one) the laser probes laterally the border of the device and by varying the position it is possible to obtain information about the collected charges or the electric field at different depths from the detector surface. Another technique is the \mbox{top-Two-Photon} Absorption, which exploits a \mbox{high-intensity} pulsed mode-locked laser to obtain time coincidence and a microfocusing system to have the space overlap essential to reach a proper energy confinement. In the right plot of Figure~\ref{fig:TCT} there is an example of TPA applied to an LGAD sample, carried out at CERN~\cite{2018Casse_VERTEX,2019Fernandez_VCI}, where the incident beam is ellipsoidal with a longitudinal dimension of the focus \mbox{$\sigma=13~\mu$m} (gaussian). Shifting it from the top of the LGAD to the bottom it is possible to extract the electric field or the collected charge, as one may see in the picture.

Detector characterization not only means measuring real devices but also designing new ones. This step is achieved mostly through the numerical simulation, like in the Technology \mbox{Computer-Aided} Design (TCAD) approach. To simulate particle detectors, besides the usual mathematical framework based on the \mbox{Drift-Diffusion} (DD) method, we also need proper \mbox{physics-based} models able to reproduce and predict the behavior of real structures, both by the electrostatic and radiation-hardness performances. To this aim, RD50 has a dedicated community of experts in modeling and simulations of processes and devices, that can transfer and use the information coming from experimental data as TCAD input parameters. Calibrating the physical models involved allows to design, develop and optimize devices, according to all the possible conditions (sensor design and material, irradiation fluence and particle type, annealing, and so on) and in close collaboration with other simulation groups (like CMS and ATLAS).

\begin{figure}[!htb]
\centering
\includegraphics[width=0.75\linewidth]{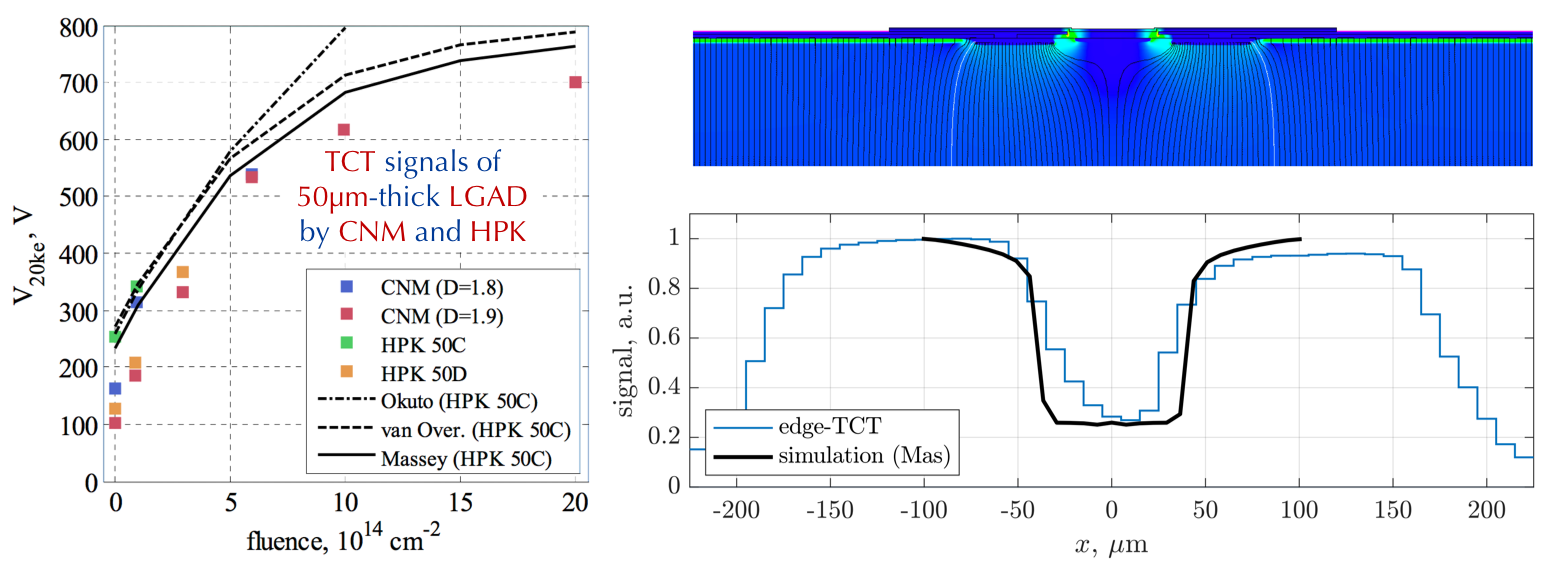}
\caption{Left: simulated (lines) and measured (symbols) bias voltage needed to collect 20 kelectrons as a function of fluence in various \mbox{50~$\mu$m~thick} LGAD by CNM and HPH. \mbox{Top-right:} field color map and drift lines as simulated in a \mbox{inter-pad} region between two LGAD. \mbox{Bottom-right}: simulated(solid black)/measured(light blue) collected charge scan versus position in the \mbox{inter-strip} region of an LGAD detector from FBK, Trento.}
\label{fig:TCAD}
\end{figure}
The main targets of the TCAD design approach are the proper modeling of the electrical behavior ($I(V,T)$ and $C(V,T)$ curves, breakdown, depletion voltage), the prediction of field/mobility, charge generation (and/or multiplication) and radiation damage effects (surface/bulk defects, trapping, acceptor \mbox{de-activation}).

Figure~\ref{fig:TCAD} shows some examples of TCAD simulations of LGAD structures~\cite{2017Mandurrino_RD50}: on the left we reported the comparison between simulated and measured voltage to have 20 kelectrons versus fluence in several \mbox{50~$\mu$m~thick} LGAD by Centro Nacional de Microelectr\'{o}nica (CNM), Spain, and Hamamatsu Photonics K.~K. (HPK), Japan. Measurements are taken with the TCT while the simulation accounts for three different charge multiplication models: the \mbox{Okuto-Crowell}, \mbox{van Overstraeten-de Man} and Massey (see Ref~\cite{2017Mandurrino_NSS}). This represents a clear example of model calibration since by comparing calculated and experimental data is possible to refine the physical input parameters. On the \mbox{top-right} picture, there is a simulated LGAD \mbox{cross-section} with the electric field color map and drift lines. Such studies allow to optimize the termination structures between neighboring active areas. Finally, the \mbox{bottom-right} plot shows measured (\mbox{edge-TCT}) and simulated (with Massey model for charge multiplication) signal scan between two adjacent LGAD, very useful to define the detector overall \mbox{fill-factor}, a critical parameter of timing trackers in future \mbox{high-luminosity} colliders.

\section{Novel structures and technologies}
\label{section4}

Besides the investigation on materials and existing detectors, the RD50 Collaboration also takes particular care of designing and developing novel structures and \mbox{full-detection} systems. Thanks to the close collaboration with the other experiments at CERN, this last R\&D program is always ongoing, and represents one of the most advanced and challenging activities, especially for what concerns the prediction of detectors performances at \mbox{very-high} fluences.

\begin{figure}[!htb]
\centering
\includegraphics[width=0.9\linewidth]{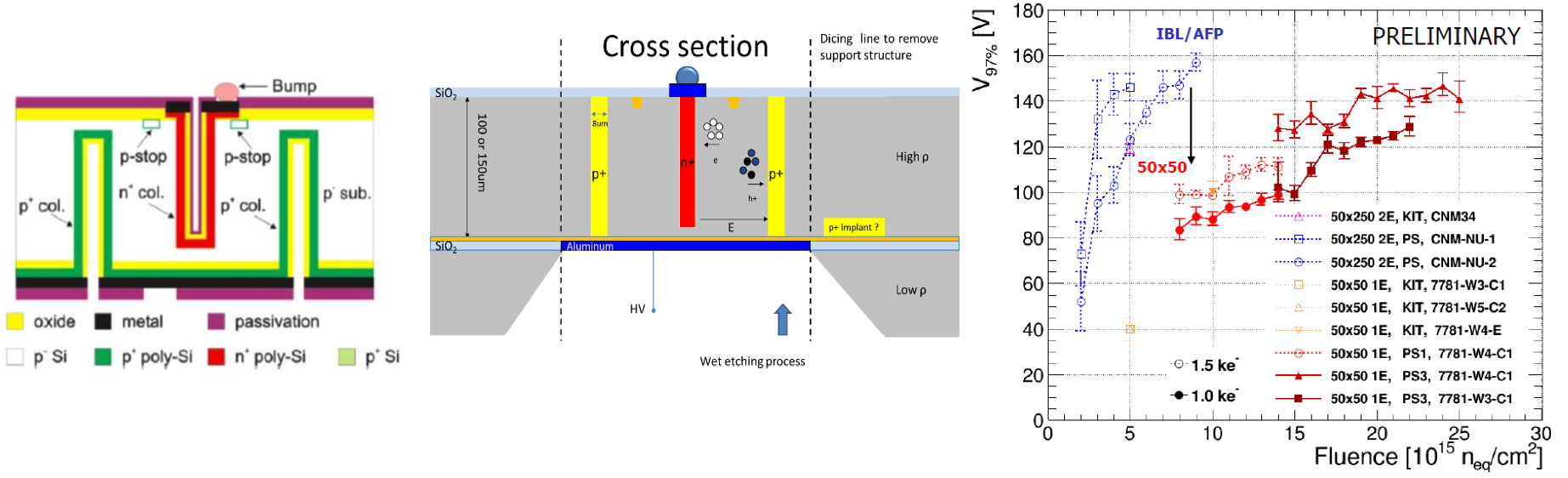}
\caption{Left-middle: \mbox{cross-sections} of a double- and \mbox{single-sided-process} 3D detector, respectively. Right: voltage needed to have 97\% efficiency as a function of fluence in several 3D detectors~\cite{2018Lange_JINST}.}
\label{fig:3D}
\end{figure}

Among the novel technologies we find the 3D and planar (\mbox{strip/pixel}) detectors. The first ones have been proposed by S. I. Parker and C. Kennedy in 1997~\cite{1997Parker_NIMA} and are based on the implementation of vertical column electrodes penetrating inside the detector bulk. Their development, nowadays carried out by two main producers (FBK and CNM), has been mainly focused on innermost pixel layers of the ATLAS experiment (especially for the Insertable \mbox{B-Layer}, IBL) plus several joint MPW productions with CMS and LHCb. Due to their design, the main detector \mbox{figures-of-merit} are a low depletion voltage (i.e., low power dissipation), charge drift decoupled from particle track (allowing short collection time), low drift path (so, reduced trapping) and reduced charge sharing. Despite their quite complicate fabrication process, that can be both double- and \mbox{single-sided}, they demonstrated noticeable efficiency even up to a fluence of \mbox{$\sim$$3\cdot10^{16}$~n$_\mathrm{eq}$/cm$^2$} (see Figure~\ref{fig:3D} and Ref.~\cite{2018Lange_JINST}) and also good time resolution: \mbox{$\sigma_t \sim 30 \,\mathrm{ps}$} at \mbox{$V_\mathrm{bias} > 100 \,\mathrm{V}$} and \mbox{$T = -20^\circ\mathrm{C}$}~\cite{2019Kramberger_TREDI}.

Other novel structures developed within the RD50 framework are the \mbox{High-Voltage} \mbox{(HV)-CMOS}, consisting in depleted active pixel detectors realized with the CMOS process. Their sensitive element are deep \mbox{$n$-wells} in (usually) low resistivity (\mbox{$\sim$10~$\Omega\cdot$cm}) \mbox{$p$-type} substrates, on handle carriers that can be thinned down. The main charge collection mechanism is the drift, and it takes place in a very thin depletion region (\mbox{$\sim$10-20~$\mu$m}). Typically a depleted thickness of \mbox{$\sim$10~$\mu$m} at \mbox{$\sim$60~V} may lead to charge collection of \mbox{$\sim$1000~electrons}. For what concerns their \mbox{radiation-related} performances, edge-TCT measurements demonstrate that, due to the trapping effect, they partly loose the diffusion component of the signal when irradiated up to \mbox{$\sim$10$^{15}$~n$_\mathrm{eq}$/cm$^2$}. Then, from 1 to some units of \mbox{10$^{15}$~n$_\mathrm{eq}$/cm$^2$} the CCE rises again even above the initial (\mbox{not-irradiated}) value, due to acceptor \mbox{de-activation}. Finally, at higher fluence, there is a second CCE degradation, induced by an increase of $N_\mathrm{eff}$ in the space charge region and the subsequent trapping~\cite{2016Affolder_JINST}. As already demonstrated in Ref.~\cite{2018Mandic_RD50}, their acceptor \mbox{de-activation} is quite similar to what observed in LGAD structures.

Before introducing the last topics of this work, we want to spend few words on one of the most interesting results in combining sensor and readout electronics, both developed within the RD50 framework. A \mbox{full-detector} system prototype for \mbox{HL-LHC} composed by coupling the RD53A chip to a 3D pixel sensor has been tested on the SPS beam at CERN (\mbox{120 GeV} \mbox{$\pi^+$/p}). Efficiency maps collected at 0$^\circ$ tilting angle demonstrated that, even after a fluence of \mbox{10$^{16}$~n$_\mathrm{eq}$/cm$^2$}, all tested samples lost only less than 1\% of their initial efficiency before irradiation~\cite{2018Alonso_RD50,2018Giannini_RD50}.

To the aim of developing planar sensors for timing applications, RD50 introduced the LGAD technology. The first samples were designed and produced by CNM in 2014~\cite{2014Pellegrini_NIMA}, and actually they are produced also by FBK and HPK, with a strong participation of the Italian INFN institute. LGAD ensure large and fast signals with a high S/N ratio, being based on the internal multiplication mechanism provided in \mbox{reverse-bias} conditions by a \mbox{$p^+$-type} layer, implanted just beneath the \mbox{$n$-cathode}. A limit of the standard LGAD technology is the way to obtain the spatial resolution required by timing applications in \mbox{high-energy} physics experiments. Being each active area surrounded by proper termination structures, ensuring electrostatic isolation and confining the high multiplication fields, such devices are characterized by the \mbox{so-called} \mbox{dead-area}, i.e. a region where there is no multiplication. These area induces a degradation of the overall efficiency of the detector, reflecting on the timing performances. To overcome this issue, RD50 is implementing various alternative solutions. Among the most interesting ones there is the use of \mbox{high-aspect-ratio} isolation trenches in place of termination implants to properly control the electric field while minimizing the impact of having a \mbox{dead-area}. A second and quite elegant solution are the ``RSD'' and ``AC-LGAD'' projects~\cite{2018Mandurrino_TREDI,2017Sadrozinski_RD50}, aiming at introducing the resistive \mbox{AC-coupled} readout paradigm, where multiplied charges are freezed in a resistive layer before being completely induced on proper capacitively coupled pads. The advantage of this technology is the use of continuous gain layers throughout the whole sensor, without the presence of any \mbox{no-gain} zone. Some samples have been already produced by CNM while a new run has been recently released by FBK in the collaboration framework with INFN Torino, Italy.

Devices based on the LGAD technology have now been chosen to be the final detectors for the timing upgrades at \mbox{HL-LHC}. But in this case the acceptor \mbox{de-activation} is particularly critical since most of the Boron inactivated comes from the multiplication layer. The result is that with the increasing of fluence the collected charge and, thus, the signal experience a significant degradation due to the gain loss. Consequently, the time resolution increases. To mitigate the \mbox{de-activation} both CNM and FBK developed LGAD with the \mbox{co-implantation} of Carbon in the multiplication layer or with the Gallium in place of Boron as a dopant atom of the \mbox{$p^+$-type} implant. If satisfactory results have been obtained with Carbon, since it seems that the interstitials complexes between Silicon and Carbon prevent the Boron \mbox{de-activation}, no significant improvements come from the Gallium~\cite{2017Ferrero_NIMA}.

The challenging question we would actually like to answer is: will our Silicon detectors still be alive beyond the final limit of \mbox{$\phi > 10^{17}$~n$_\mathrm{eq}$/cm$^2$}? Some evidences say that \mbox{very-high-fluences} mean increase of trapping up to a certain radiation level, after that the traps production saturates and their concentration remains almost constant with fluence. To get rid of the high density of defects, one could in principle make even more thin devices. But the drawback is linked to the lower signals. Some possible solutions are, (\emph{i})~again, the inclusion of a multiplication layer (thin LGAD), or (\emph{ii}) decouple drift path and total charge deposition (3D detectors). But the radiation up to \mbox{$7 \cdot 10^{17}$~n$_\mathrm{eq}$/cm$^2$} (as expected in FCC) still remains challenging and imposes new material/detectors characterizations and models.








\end{document}

%% file: econfmacros.tex
\def\beq{\begin{equation}}
\def\eeq#1{\label{#1}\end{equation}}
\def\eeqn{\end{equation}}

\def\beqa{\begin{eqnarray}}
\def\eeqa#1{\label{#1}\end{eqnarray}}
\def\eeqan{\end{eqnarray}}

\let\bar=\overbar

\def\Dslash{\not{\hbox{\kern-4pt $D$}}}
\def\dslash{\not{\hbox{\kern-2pt $\del$}}}

\def\msb{{\bar{\ssstyle M \kern -1pt S}}}

